\documentclass[aps,preprint]{revtex4}
\usepackage{graphicx}
\usepackage{amssymb}
\usepackage{amsmath}
\usepackage{color}
\usepackage{enumerate}
\usepackage{bm}
\begin{document}

%\title{Relaxation in Assembly of Dipolar Interacting Nanoparticles}
%\title{Relaxation in Ordered Assembly of Dipolar Coupled Magnetic Nanoparticles}
\title{Relaxation in Ordered Assembly of Magnetic Nanoparticles}
\author{Manish Anand}
\email{itsanand121@gmail.com}
\affiliation{Department of Physics, Bihar National College, Patna University, Patna-800004, India.}

\date{\today}

\begin{abstract}
We study the relaxation characteristics in the two-dimensional ($l^{}_x \times l^{}_y$) array of magnetic nanoparticles (MNPs) as a function of  aspect ratio $A^{}_r=l^{}_y/l^{}_x$, dipolar interaction strength $h^{}_d$ and anisotropy axis orientation using computer simulation. The anisotropy axes of all the MNPs are assumed to have the same direction, $\alpha$ being the orientational angle. 
%A common anisotropy axis assumption permits us to probe the precise role of dipolar interaction on the relaxation mechanism. 
Irrespective of $\alpha$ and $A^{}_r$, the functional form of the magnetization-decay curve is perfectly exponentially decaying with $h^{}_d\leq0.2$. There exists a transition in relaxation behaviour at $h^{}_d\approx0.4$; magnetization relaxes slowly for $\alpha\leq45^\circ$; it relaxes rapildy with $\alpha>45^\circ$.
%On the other hand, there is a fastening in the magnetization relaxation with $\alpha>45^\circ$. 
Interestingly, it decays rapidly for $h^{}_d>0.6$, irrespective of $\alpha$. It is because the dipolar interaction promotes antiferromagnetic coupling in such cases. There is a strong effect of $\alpha$ on the magnetic relaxation in the highly anisotropic system ($A^{}_r\geq25$). Interesting physics unfolds in the case of a huge aspect ratio $A^{}_r=400$. There is a rapid decay of magnetization with $\alpha$, even for weakly interacting MNPs. Remarkably, magnetization does not relax even with a moderate value of $h^{}_d=0.4$ and $\alpha=0^\circ$ because of  ferromagnetic coupling dominance. Surprisingly, there is a complete magnetization reversal from saturation (+1) to $-1$ state with $\alpha>60^\circ$. The dipolar field and anisotropy axis tend to get aligned antiparallel to each other in such a case. The effective N\'eel relaxation time $\tau^{}_N$ depends weakly on $\alpha$ for small $h^{}_d$ and $A^{}_r\leq25.0$. For large $A^{}_r$, there is a rapid fall in $\tau^{}_N$ as $\alpha$ is incremented from 0 to $90^\circ$.
These results benefit applications in data and energy storages where such controlled magnetization alignment and desired structural anisotropy are desirable.

\end{abstract}

\maketitle

\section{Introduction}
In recent years, there has been much research interest in two-dimensional %self-assembled
arrays of magnetic nanoparticles (MNPs) due to their diverse technological applications such as spintronics, magnetic hyperthermia, drug delivery, data storage, etc.~\cite{leo2018,puntes2004,mohammadpour2020,farhan2013,kechrakos2002,wang2008,bupathy2019}. In these contexts, relaxation characteristics of the underlying system are one of the essential quantifiers~\cite{ota2019,waintal2003}.
%the understanding of magnetic relaxation is of crucial importance. 
The latter is primarily characterized by a time scale known as N\'eel relaxation time, which depends strongly on various parameters of interest such as particle size, thermal fluctuations, anisotropy strength, magnetic interaction, etc.~\cite{fabris2019,hergt2009,tackett2015}. Therefore, the investigation of magnetic relaxation in such a system represents a topic of practical importance.

%The thermal fluctuations and relaxation of
%the magnetization of single-domain particles currently merit
%attention in the context of information storage and rock mag-
%netism.

The magnetic relaxation properties are well understood in the case of non-interacting MNPs~\cite{wernsdorfer1997,carrey2011}. However, MNPs are found to interact primarily via dipole-dipole interaction in an assembly.
%it plays a crucial role in determining the phase behavior and morphology of the magnetic ordering.
%As the dipolar interaction is long-ranged and anisotropic in nature, 
The dipolar interaction has varied effects on various thermodynamical and magnetic properties of crucial importance because of its long-range and anisotropic behaviour~\cite{anand2016,anand2018}. For instance, it imparts spin-glass like character in randomly distributed MNPs~\cite{konwar2020,parker2008}. On the other hand, it plays a crucial role in determining the 
morphology of magnetic ordering ~\cite{santos2020,morup2010}. Holden {\it et al.} studied the ground state spin structures in two-dimensional kagome lattice using Monte Carlo simulation. They observed six-fold degenerate spin states because of dipolar interaction~\cite{holden2015}. Bailly-Reyre {\it et al.} found ground-state configuration to be a vortex in the cubic assembly of nanodots~\cite{bailly2021}. Luttinger {\it et al.} observed the minimum energy configurations to be ferromagnetic in a face-centred cubic lattice~\cite{luttinger1946}. In contrast, it is antiferromagnetic in a simple cubic arrangement of MNPs~\cite{luttinger1946}. It promotes antiferromagnetic spin states in a square array, while for a triangular arrangement, the minimum energy configuration is ferromagnetic~\cite{macisaac1996,politi2006}. The dipolar interaction may induce ferromagnetic or antiferromagnetic coupling among the MNPs depending on their relative positions. Consequently, the magnetic relaxation properties of interacting MNPs not only depends on interaction strength but also on the spatial configuration of particles~\cite{dejardin2011}. %In addition to  
%because of anisotropic and long-ranged nature of dipolar interaction~\cite{dejardin2011}.

Various research works suggest that the dipolar interaction affects the relaxation characteristics in an assembly of MNPs strongly~\cite{iglesias2004,liao,anand2021ther}. For example, Gallina {\it et al.} theoretically investigated the magnetization dynamics and interaction energy landscapes in a two-dimensional assembly of dipolar interacting MNPs~\cite{gallina2020}. Magnetic relaxation is found to follow stretched exponential law in weakly disordered systems. Using computer simulations, Patrick Ilg studied the relaxation dynamics of multicore magnetic nanoparticles~\cite{ilg2017}. The magnetic relaxation is well characterized by an exponentially decaying function for moderate dipolar interaction strength. Denisov {\it et al.} probed the relaxation properties in two-dimensional assembly using mean-field approximations~\cite{denisov2002}. They observed a two-distinct relaxation time scale. Shtrikmann {\it et al.} and Dormann {\it et al.} investigated the magnetic relaxation in an interacting assembly using theoretical calculations~\cite{shtrikman1981,dormann1988}. They also observed an elevation in relaxation time due to dipolar interaction. On the other hand, M{\o}rup {\it et al.} observed a decrease in relaxation time with an increase in dipolar interaction strength~\cite{morup1994}. In recent work, we studied magnetic relaxation in the two-dimensional assembly of MNPs as a function of  dipolar interaction strength and aspect ratio of the system with randomly oriented anisotropy axes~\cite{anand2021}. The dipolar interaction of enough strength increases or decreases the relaxation time depending on the aspect ratio of the system.

Some recent works also indicate that the orientation of anisotropy axes plays a crucial role in determining various magnetic properties of interest~\cite{valdes2020,jiang2016,anand2020}. It is also strengthened from the fact that the magnetic field promotes some degree of orientation of elongated structures along the field direction~\cite{martinez2013,serantes2014,conde2015}. Conde-Leborán {\it et al.} studied the heating efficiency in the assembly of interacting MNPs as a function of the degree of collinearity of their easy axes~\cite{conde2015}. The amount of heat dissipation depends strongly on the anisotropy axes orientation for weakly interacting MNPs. Boekelheide {\it et al.} investigated the effect of anisotropy axes orientation on the hysteresis loops using micromagnetic simulations and experiments~\cite{boekelheide2019}. 
The coercivity is found to significantly modified by the direction of anisotropy axes. Using experiments, Li {\it et al.} studied the hysteresis response in dense arrays of magnetic nanowires~\cite{li2020}. The coercive field and remanent magnetization are more significant with perfectly aligned anisotropy as compared to perpendicular orientation.
%Hysteresis curves reveal that values of magnetic coercivity and remanent magnetization in the preferred magnetization direction are both higher than that of the nanoparticles, while these values in the perpendicular direction are both lower.
Allia {\it et al.} analyzed the hysteresis properties in an assembly of MNPs with aligned and randomly oriented anisotropy axes~\cite{allia2020}. The hysteresis loop area is enormous with aligned anisotropy axes as compared to the random orientation case. Using kinetic Monte Carlo simulations and analytical calculations, we studied the magnetic relaxation in the linear chain of nanoparticles as a function of anisotropy axis orientation and dipolar interaction strength~\cite{anand2019}. There is a fastening or slowing down of magnetization relaxation depending on the 
angle between the anisotropy axis and the particles array.

It is evident from above discussions that the dipolar interaction and anisotropy axes orientation strongly affect the magnetic relaxation characteristics in ordered arrays of MNPs. However, a complete understanding of the effect of these factors is still lacking. Thus motivated, we systematically analyzed the effect of dipolar interaction, the orientation of anisotropy axes, and the aspect ratio of the system on the relaxation mechanism in two-dimensional assembly of MNPs using kinetic Monte Carlo (kMC) simulation in the present work. We also investigate the variation of effective N\'eel relaxation time $\tau^{}_N$ as a function of these parameters.   

%However, a complete model able to account for the observed dynamic properties of isolated as well as magnetically interacting particles is still lacking.
The rest of the paper is organized as follows: We discuss the model and various energy terms in Sec.~II. The simulation method is also discussed in  brief. The numerical results are analyzed in Sec.~III. Finally, we provide the summary of the present work in Sec.~IV.

%In Sec.~II, we present the model and discuss the various energy terms. We also discuss the kMC simulations briefly. 

%The simulation results will be discussed in Sec.~III. Finally, we summarize and conclude the work in Sec.~IV.

%It is evident from above that the effect of dipolar interaction on magnetic relaxation is not completely understood despite numerous work. Evaluation of $\tau_N$ is also crucial because of its important role in various technological applications [53–57]. For instance, Kuncser et al. observed that the amount of heating effciency of MNPs depends strongly on $\tau_N$ [55]. In a recent study, we have also shown that relaxation time is dictated by the dipolar interaction in a linear arrangement of MNPs, one of the essential quantifiers in determining the heat dissipation in such a system [56]. $\tau_N$ also plays a significant role in data storage applications [57]. Therefore, we systematically investigate the role of dipolar interaction, temperature and aspect ratio on the magnetic relaxation in a two-dimensional assembly of superparamagnetic nanoparticle using state of the art kinetic Monte Carlo simulation (kMC).

\section{Model}
\begin{figure}[!htb]
	\centering\includegraphics[scale=0.50]{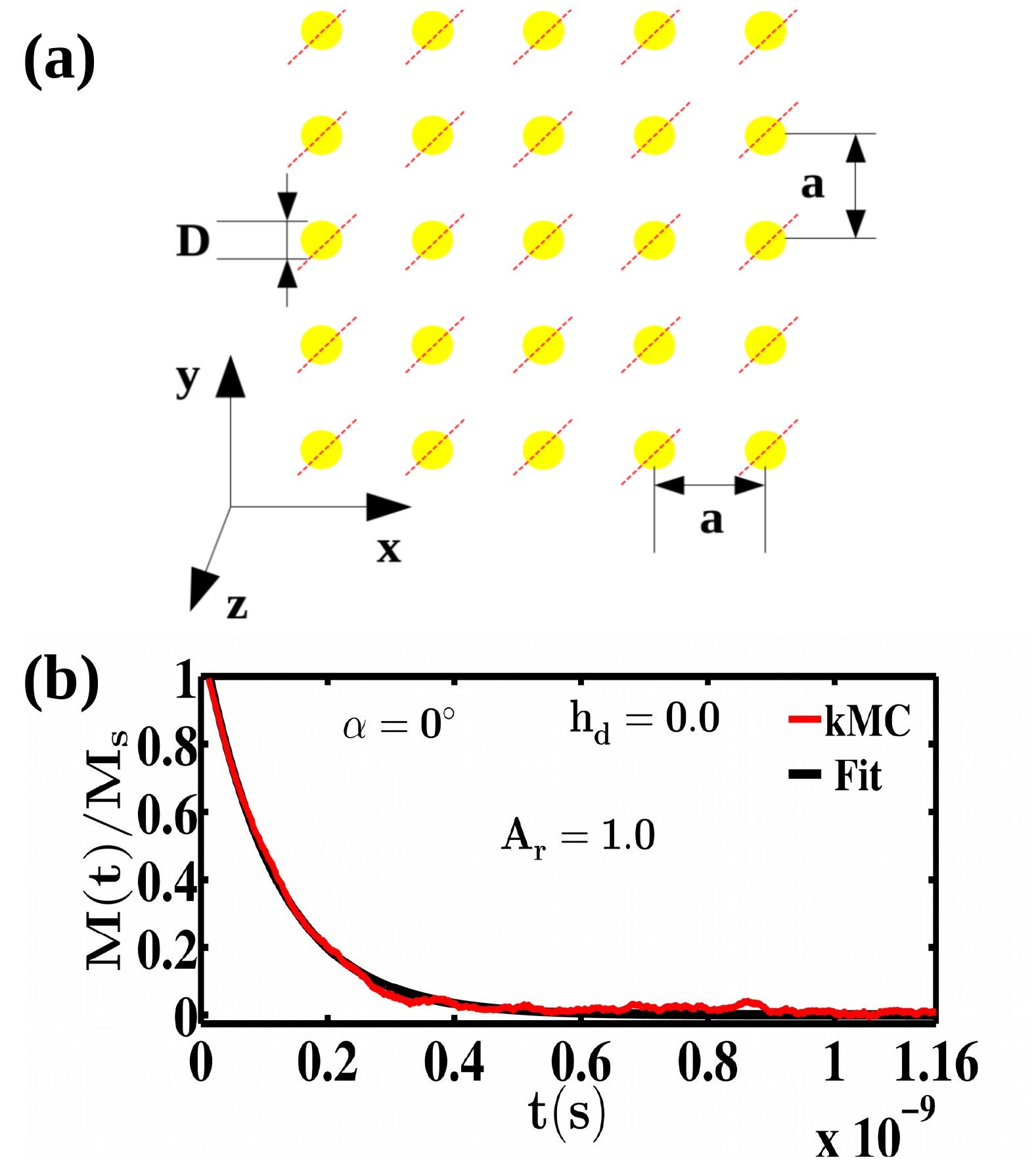}
	\caption{(a) Schematic of the two-dimensional array of magnetic nanoparticles. The diameter of the nanoparticle is $D$, and $a$ is the lattice spacing. The dashed line denotes the direction of the anisotropy axis. All the particles are assumed to have the same anisotropy axes orientations, $\alpha$ being the angle with respect to the $y$-axis. (b) Magnetization-decay $M(t)/M^{}_s$ vs. $t$ curve for non-interacting MNPs ($h^{}_d=0.0$) with $A^{}_r=1.0$ and perfectly aligned anisotropy axes ($\alpha=0^\circ$). We have fitted the simulated curve with $M(t)/M^{}_s=\exp(-t/\tau^{o}_N)$ and shown it with a black line. In the absence of magnetic interaction, the form of the relaxation curve does not depend on $A^{}_r$ and $\alpha$ (curves not shown).}
	\label{figure1}
\end{figure}

We consider $N$ number of spherical shaped nanoparticles arranged on $l^{}_x\times l^{}_y$ two-dimensional lattice. Let the lattice constant be $a$ and the particle diameter be $D$, as depicted in Fig.~\ref{figure1}(a). Each nanoparticle has a magnetic moment $\mu=M^{}_sV$, $M^{}_s$ being the saturation magnetization and $V=\pi D^3/6$ is the MNP volume. The anisotropy axes of all the MNPs are assumed to have the same orientation with respect to the $y$-axis of the sample as shown in the schematic Fig.~\ref{figure1}(a), $\alpha$ is the orientational angle. 
%We consider such a situation because for two main reasons: 
The choice of aligned anisotropy axis is considered for two
reasons: (1) The assembly with aligned anisotropy axes can be realized experimentally~\cite{deng2020,wen2017,jiang2020}. They have distinct magnetic features, which could be useful in various applications.  (2) It also provides a unique pathway to distinctly analyze the role of dipolar interaction and anisotropy on the relaxation mechanism.  
%We are also able to probe the precise role of dipolar interaction.

%The energy of a single MNP due to magnetocrystalline anisotropy is given by the following relation
The following relation gives the energy of a single MNP due to magnetocrystalline anisotropy~\cite{anand2019,muscas2018} 
\begin{equation}
E^{}_K=K^{}_\mathrm {eff} V\sin^2\theta.
\label{barrier}
\end{equation}
Here $\theta$ is the angle between the anisotropy axis or the easy axis and the magnetic moment. In the absence of interaction, the functional form of Eq.~(\ref{barrier})  is a symmetric double-well having two energy minima at $\theta=0$ and $\pi$, respectively. An energy maximum of strength $K^{}_\mathrm {eff}V$ at $\theta=\pi/2$ separates these minima, also termed as energy barrier. There is a finite probability for the magnetic moment to flip and reverse its direction in the presence of sufficient temperature. The mean time between two flips is known as the N\'eel relaxation time $\tau^{o}_{N}$  and is given by the N\'eel-Arrhenius equation~\cite{anand2019,carrey2011}

%between the magnetic moment and the easy axis. The functional form of the single-particle energy function [Eq.~(\ref{barrier})] is a symmetricdouble well. This energy function has two energy minima at $\theta=0$ and $\pi$, respectively. There is a energy maximum or the energy barrier of strength $K^{}_\mathrm {eff} V$ at $\theta=\pi/2$ as depicted in Fig.~\ref{figure1}(b). 
%In the presence of thermal fluctuations, the magnetic moment has a finite probability of flipping and reversing its direction by overcoming the energy barrier $K^{}_\mathrm {eff} V$. The mean time between two flips is termed the Néel relaxation time  $\tau^{o}_{N}$ defined as~\cite{anand2019,carrey2011}
\begin{equation}
\tau^{o}_{N}=\tau^{}_o\exp(K^{}_\mathrm {eff}V/k^{}_BT).
\label{Neel}
\end{equation}
Here $\tau^{}_o=(2\nu_o)^{-1}$, $\nu_o\approx10^{10}$ $s^{-1}$ is the attempt frequency. $T$ is the temperature, and $k^{}_B$ is the Boltzmann constant. Eq.~(\ref{Neel}) is applicable for a single nanoparticle or very dilute assembly of MNPs.

%applicable for non-interacting MNPs.

In an assembly, magnetic nanoparticles primarily interact because of dipolar interaction. The energy associated with such interaction can be evaluated using the following expression~\cite{usov2017,anand2021hys}

%In an assembly, MNPs primarily interact due to long-ranged dipolar interaction. We can calculate the dipolar interaction energy $E^{}_{\mathrm {dip}}$ in such a case as~\cite{odenbach2002,usov2017,anand2021hys}
%\begin{equation}
%\label{dipole}
%E_{dd}(s)= -\frac{1}{4\pi \mu_{o}}\left(\frac{3\left(\vec{\mu_{i}}\cdot\vec{s}\right)\left(\vec{\mu_{j}}\cdot \vec{s}\right)}{s^{5}} -\frac{\vec{\mu_{i}}\cdot \vec{\mu_{j}} }{s^3}\right),
%\end{equation}
\begin{equation}
\label{dipole}
E^{}_{\mathrm {dip}}=\frac{\mu^{}_o\mu^2}{4\pi a^3}\sum_{j,\ j\neq i}\left[ \frac{\hat{\mu_{i}}\cdot\hat{\mu_{j}}-3\left(\hat{\mu_{i}}\cdot\hat{r}_{ij}\right)\left(\hat{\mu_{j}}\cdot\hat{r}_{ij}\right)}{(r^{}_{ij}/a)^3}\right].
\end{equation}
Here $\mu_{o}$ is the permeability of free space; $\hat{\mu}_{i}$ and $\hat{\mu}_{j}$ are the unit vectors for the magnetic moment of $i^{th}$ and $j^{th}$ nanoparticle, respectively, and the center-to-center separation between them is $r^{}_{ij}$, $\hat{r}^{}_{ij}$ is the corresponding unit vector. 
%is the center-to-center separation between them.  $\hat{r}^{}_{ij}$ is the unit vector corresponding to $\vec{r}_{ij}$.
%The particle has a magnetic moment $\mu=M^{}_sV$, $M^{}_s$ is the saturation magnetization. 
%$\mu^{}_o\vec{H}^{}_{\mathrm {dip}}$ is the dipolar field due to all other nanoparticles, present in the system. 
%The corrspoding field  

We can calculate the dipolar field $\mu^{}_o\vec{H}^{}_{\mathrm {dip}}$ corresponding to the dipolar interaction as~\cite{anand2021hys,tan2014}
\begin{equation}
\mu^{}_{o}\vec{H}^{}_{\mathrm {dip}}=\frac{\mu\mu_{o}}{4\pi a^3}\sum_{j,j\neq i}\frac{3(\hat{\mu}^{}_j \cdot \hat{r}_{ij})\hat{r}^{}_{ij}-\hat{\mu^{}_j} }{(r_{ij}/a)^3}.
\label{dipolar1}
\end{equation}
%Here $\mu_{j}$ is the magnetic moment vector of the $j^{th}$ particle, $s$ is the center-to-center separation between $\mu_{i}$ and $\mu_{j}$; and the $\mu_{o}$ is the permeability of free space.
Eq.~(\ref{dipole}) and Eq.~(\ref{dipolar1}) clearly suggest that the strength of this long-ranged interaction varies as $1/r^{3}_{ij}$. Therefore, we can define a parameter $h^{}_d=D^{3}/a^3$~\cite{tan2010} to model the dipolar interaction strength. As $h^{}_d= 1.0$ implies $D=a$, the separation between the two nearest neighbouring MNP is the least. Consequently, the dipolar interaction strength is the maximum in this case. Likewise, $h^{}_d=0$ mimics the non-interacting state. We can write the total energy of the system as~\cite{tan2014,anand2019}
\begin{equation}
E=K^{}_{\mathrm {eff}}V\sum_{i}\sin^2 \theta^{}_i+\frac{\mu^{}_o\mu^2}{4\pi a^3}\sum_{j,\ j\neq i}\left[ \frac{\hat{\mu_{i}}\cdot\hat{\mu_{j}}-{3\left(\hat{\mu_{i}}\cdot\hat{r}_{ij}\right)\left(\hat{\mu_{j}}\cdot\hat{r}_{ij}\right)}}{(r_{ij}/a)^3}\right]
\end{equation}
%Here $\theta^{}_i$ is the angle between the anisotropy axis and the $i^{th}$ magnetic moment of the system. 

It is clearly evident from the above discussion that the single-particle energy function Eq.~(\ref{barrier}) gets altered because of dipolar interaction. Consequently, the modified energy function has new energy extrema. Let these energy minima be $E^{}_1$ and $E^{}_2$ and maxima $E^{}_3$. In the presence of thermal fluctuations, the magnetic moment tends to change its orientation.  
%The dipolar interaction modifies the single-particle energy barrier given by Eq.~(\ref{barrier}) is modified due to the dipolar interaction. Consequently, the single-particle energy function defined  becomes asymmetric, as depicted in Fig.~\ref{figure1}(c). 
%The modified energy function
%The modified energy function has single energy minimum when the dipolar field is larger than the anisotropy field $H^{}_K=2K^{}_{\mathrm {eff}}/M^{}_s$~\cite{carrey2011}.
%two new energy minima $E^{}_1$ and $E^{}_2$, and a maxima $E^{}_3$ for  $|\mu^{}_{o}\vec{H}^{}_{\mathrm {dip}}|<H^{}_K$, as shown in Fig.~\ref{figure1}(c).
Therefore, the rate $\nu^{}_1$ at which the magnetic moment goes from $E^{}_1$ to $E^{}_2$ via $E^{}_3$ can be expressed as~\cite{hanggi1990}
\begin{equation}
\nu^{}_1=\nu^{0}_{1}\exp\bigg(-\frac{E^{}_3-E^{}_1}{k^{}_BT}\bigg)
\end{equation}
Similarly, the jump rate $\nu^{}_2$ for the magnetic moment to switch its direction from $E^{}_2$ to $E^{}_1$ is given by~\cite{hanggi1990}
\begin{equation}
\nu^{}_2=\nu^{0}_2\exp\bigg(-\frac{E^{}_3-E^{}_2}{k^{}_BT}\bigg),
\end{equation} 
Here $\nu^{0}_{1}=\nu^{0}_{2}=\nu^{}_{o}$. 

We have used the kinetic Monte Carlo simulation technique to analyze magnetic relaxation. We have used the same algorithm in the present work, described in detail in the references~\cite{anand2019,tan2014,anand2021}. Therefore, we do not reiterate it to avoid repetitions. In this procedure, we first saturate all the magnetic moments along the $y$-direction of the system by applying a huge magnetic field of strength $\mu_oH_{\mathrm{o}} =20$ T. Next, we divide the total simulation time into  2000 equal steps and switch off the external field $\mu_oH_{\mathrm{o}}$ at $t=0$ s. We then study the time evolution of magnetization of the underlying system using the kMC simulation. Finally, we fit the so-obtained  magnetization-decay curve with $M(t) = M^{}_s \exp(-t/\tau^{}_N)$ to extract effective N\'eel relaxation time $\tau^{}_N$ of the underlying system.
%We have used kMC simulation to probe the magnetic relaxation and also to evaluated time $\tau^{}_N$. 

%For this, we apply a very large external magnetic field $\mu_oH_{\mathrm{max}}=20$ Tesla along the $y$-direction with respect to the underlying system so that all the moments get aligned along the external field direction. We divide the total simulation time into 2000 equal time steps, and switch off $\mu_oH_{\mathrm{max}}$ at time $t=0$ to study the time dependence of magnetic relaxation. 

%\newpage
\section{Simulations Results}
We consider spherical nanoparticles of magnetite (Fe$_3$O$_4$) with the following values of system parameters: $D=8$ nm, $K_{\mathrm {eff}}=13\times10^3$ Jm$^{-3}$, $M^{}_s=4.77\times10^5$ Am$^{-1}$, and $T=300$ K. The total number of MNPs considered as $N=400$. We have considered six values of system sizes viz. $l_x\times l_y=20\times20$, $10\times40$, $8\times50$, $4\times100$, $2\times200$ and $1\times400$. The corresponding aspect ratio $A^{}_r(=l^{}_y/l^{}_x)$ of the underlying system is $1.0$, 4.0, 6.25, 25, 100 and 400, respectively. The dipolar interaction strength $h^{}_d$ is varied from 0 to 1.0. We varied the anisotropy axis orientation angle $\alpha$ between 0 to $90^\circ$.

%All the numerical data obtained have been averaged over several independent runs to obtain good statistical averaging. 

To validate the kMC method implemented in the present work, we first probe the relaxation characteristics without any magnetic interaction. In Fig.~\ref{figure1}(b), we plot the simulated magnetization-decay $M(t)/M^{}_s$ versus $t$ curve of a square array of MNPs ($l^{}_x\times l^{}_y=20\times20$, $A^{}_r=1.0$) with $h^{}_d=0.0$ and perfectly aligned anisotropy, i.e. $\alpha=0^\circ$. The functional form of the magnetization decay curve is exponentially decaying. We fit the simulated curve with $M(t)/M^{}_s=\exp(-t/\tau^{o}_{N})$, which yields $\tau^{o}_N=1.164\times10^{-10}\pm 1.25\times10^{-11}$ s. The theoretical value of $\tau^{o}_N$ [using Eq.~(\ref{Neel})] comes out to be $1.160\times10^{-10}$ s, which shows perfect agreement with the simulation and also authenticates the kMC procedure used. In the absence of dipolar interaction, the magnetic relaxation curve is independent of $A^{}_r$ and $\alpha$. Therefore, the corresponding curves are not shown to avoid duplication. 

%In the absence of magnetic interaction, the magnetization decay curve is also found to be independent of the aspect ratio $A^{}_r$ as expected. Therefore, we have not shown the corresponding curves to avoid duplication

\begin{figure}[!htb]
	\centering\includegraphics[scale=0.40]{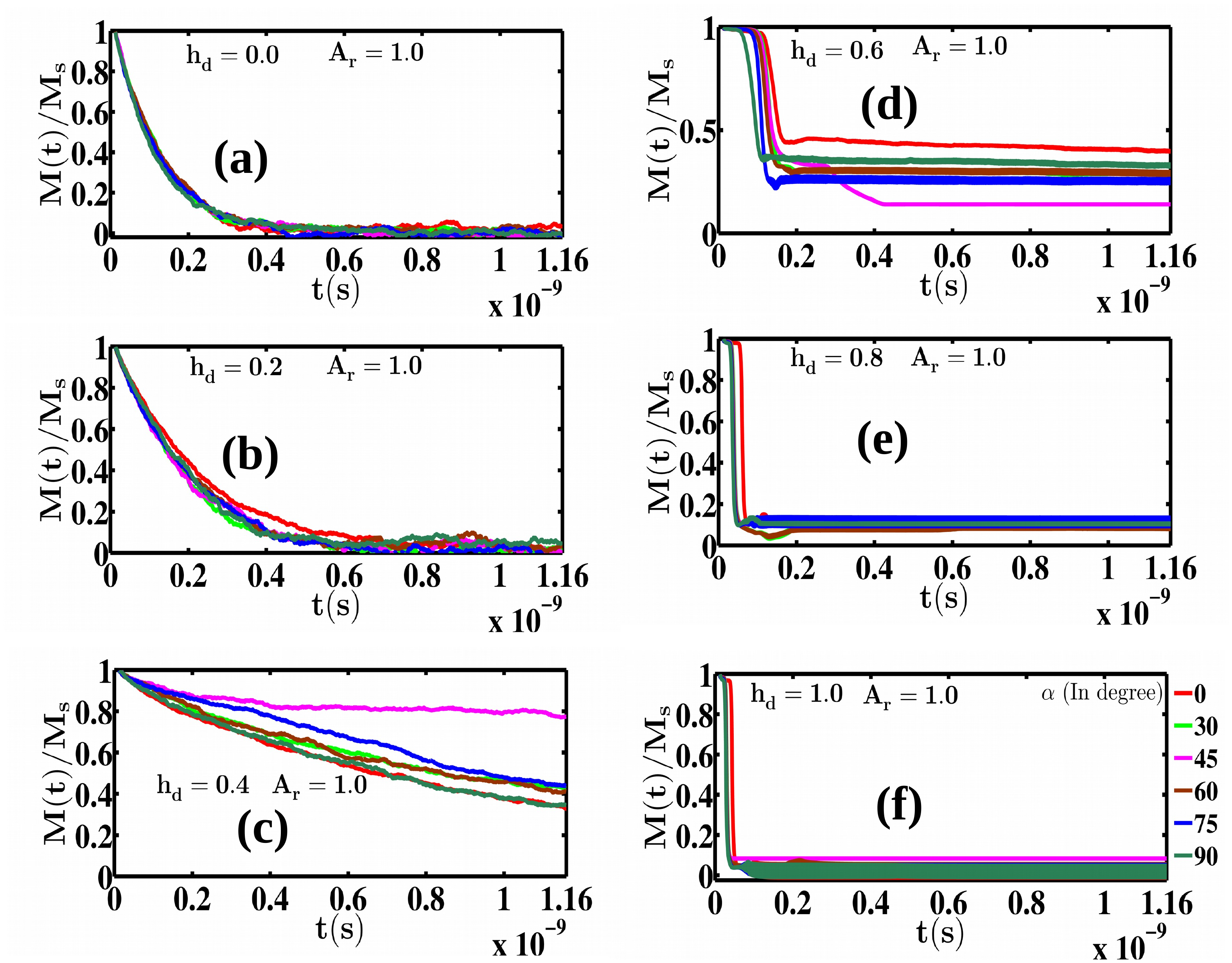}
	\caption{Magnetization-decay $M(t)/M^{}_s$ versus $t$ curves as a function of anisotropy axes orientation angle $\alpha$ with the square arrangement of MNPs ($A^{}_r=1.0$). We have considered six representative values of $h^{}_d$: (a) $h^{}_d= 0.0$, (b) $h^{}_d= 0.2$, (c) $h^{}_d= 0.4$, (d) $h^{}_d= 0.6$, (e) $h^{}_d= 0.8$, and (f) $h^{}_d= 1.0$. In the presence of small $h^{}_d\leq0.3$, the magnetization-decay curve is perfectly exponentially decaying. Interestingly, magnetization decays faster for large $h^{}_d$ in comparison with weakly interacting MNPs. There is also a weak dependence of the relaxation characteristics on $\alpha$ for a fixed $h^{}_d$.}
	\label{figure2}
\end{figure}

Next, we study the dipolar interaction and anisotropy axis orientation effect on the magnetic relaxation in a square assembly of MNPs. In Fig.~(\ref{figure2}), we plot $M(t)/M^{}_s$ versus $t$ curve with $A^{}_r=1.0$ for six typical values of $\alpha=0, 30, 45, 60, 75$, and $90^\circ$. We have also considered six representative values of $h^{}_d=0.0$, 0.2, 0.4, 0.6, 0.8, and 1.0. Irrespective of $\alpha$, the functional form of the magnetization-decay curve is perfectly exponentially decaying for weak dipolar interaction $h^{}_d\leq0.3$. There exists a transition point at $h^{}_d\approx0.4$; magnetization relaxes slowly for $\alpha\leq45^\circ$. While with $\alpha>45^\circ$, magnetization decays rapidly. Remarkably, there is a fastening in magnetization relaxation with large dipolar interaction strength ($h^{}_d>0.4$) compared to weakly interacting MNPs. It can be attributed to enhanced antiferromagnetic coupling because of dipolar interaction in the square arrangement of MNPs. Interestingly, the relaxation characteristics depend very weakly on the orientation of anisotropy axes for a given interaction strength. It could be due to the symmetric nature of the system. Figueiredo {\it et al.}~ also observed exponential decay of magnetization for weakly interacting MNPs~\cite{figueiredo2007}. The observation of fastening of magnetic relaxation due to antiferromagnetic coupling induced by dipolar interaction is in perfect agreement with our recent work~\cite{anand2021hys}. We found characteristic magnetic hysteresis of antiferromagnetic dominance in a square arrangement of MNPs~\cite{anand2021hys}. De'Bell {\it et al.} also obtained the minimum energy state to be antiferromagnetic in the square array~\cite{de1997}.

\begin{figure}[!htb]
\centering\includegraphics[scale=0.40]{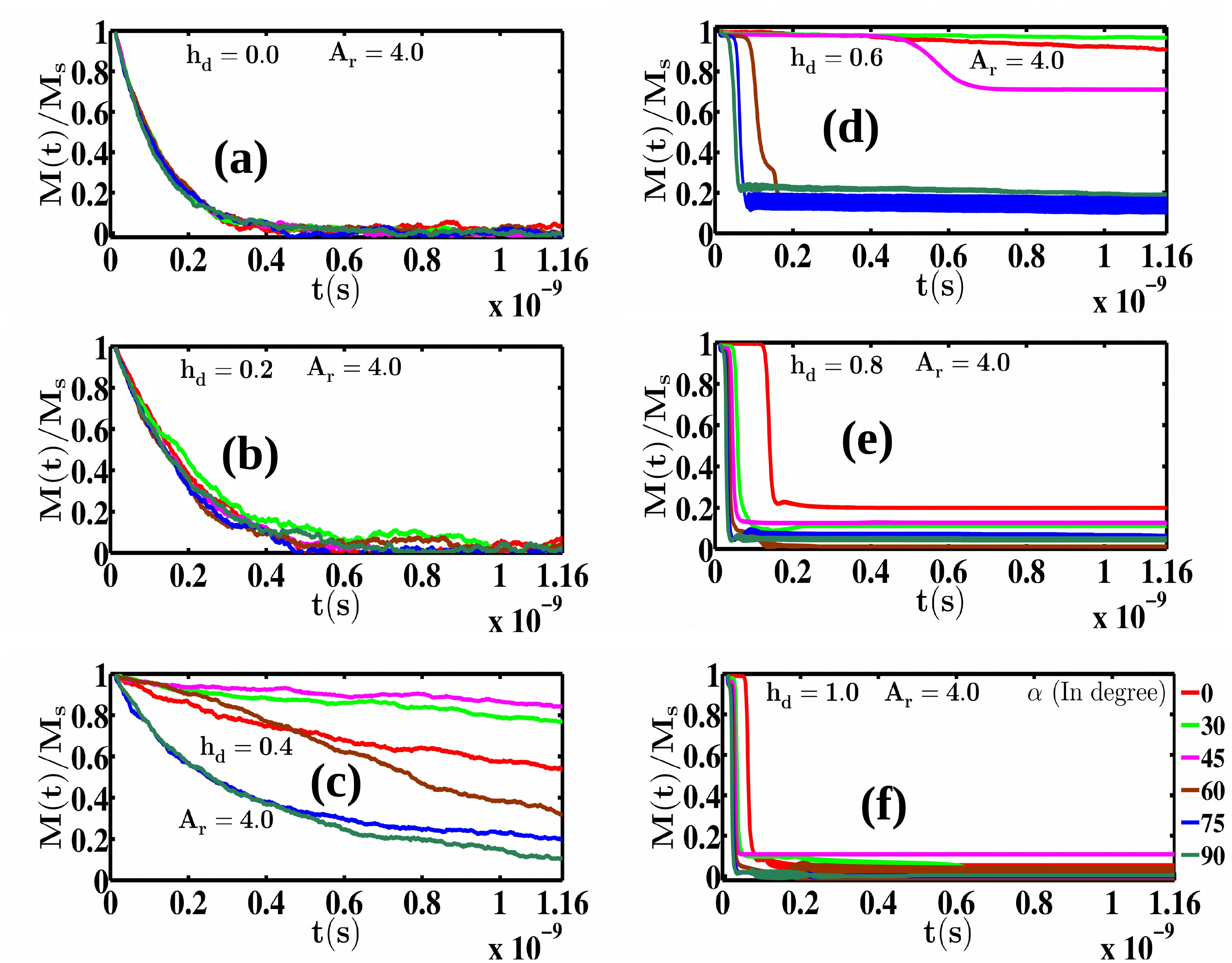}
\caption{Time evolution of magnetization as a function of $\alpha$ for various values of $h^{}_d$ with $A^{}_r=4.0$. We have considered six representative values of $h^{}_d$: (a) $h^{}_d= 0.0$, (b) $h^{}_d= 0.2$, (c) $h^{}_d= 0.4$, (d) $h^{}_d= 0.6$, (e) $h^{}_d= 0.8$, and (f) $h^{}_d= 1.0$. The magnetization decays slowly for $\alpha\leq45^\circ$ with moderate dipolar interaction strength $h^{}_d$. On the other hand, there is rapid decay of magnetization with $\alpha>45^\circ$. Irrespective of $\alpha$, magnetization decays rapidly for large $h^{}_d$ as compared to weakly dipolar interacting case.} 
\label{figure3}
\end{figure}

The easy axes orientation should affect the relaxation characteristics in an anisotropic system ($A^{}_r\neq1$). Therefore, we now analyze the time evolution of magnetization with $A^{}_r=4.0$. In Fig.~(\ref{figure3}), we plot the magnetization-decay $M(t)/M^{}_s$ vs. $t$ for $A^{}_r=4.0$ and six values of $\alpha=0, 30, 45, 60, 75$, and $90^\circ$. All other parameters are the same as that of Fig.~(\ref{figure2}). The magnetization relaxation curve is perfectly exponentially decaying for the small dipolar interaction strength ($h^{}_d\leq0.2$), similar to that of $A^{}_r=1.0$. In the presence of moderate dipolar interaction ($h^{}_d\approx0.4$), the magnetization-decay gets slower for $\alpha\leq45^\circ$. On the other hand, there is a fastening in the magnetic relaxation with $\alpha>45^\circ$. In the presence of large dipolar interaction strength, the anisotropy axes orientation affects the relaxation characteristics strongly as anticipated. The magnetization ceases to relax for $\alpha\leq30^\circ$ with $h^{}_d\approx0.6$; it relaxes faster for $\alpha>30^\circ$. The decay of magnetization is extremely rapid for $h^{}_d>0.6$, irrespective of $\alpha$. It is because the strength of antiferromagnetic coupling induced by the dipolar interaction is the maximum in these cases. {\color{black} These results clearly indicate that we can manipulate the nature of the relaxation (slowing or fastening) in a more controlled way by varying $h^{}_d$ and $\alpha$, which is an essential quantifier in spintronics based applications}.

\begin{figure}[!htb]
\centering\includegraphics[scale=0.40]{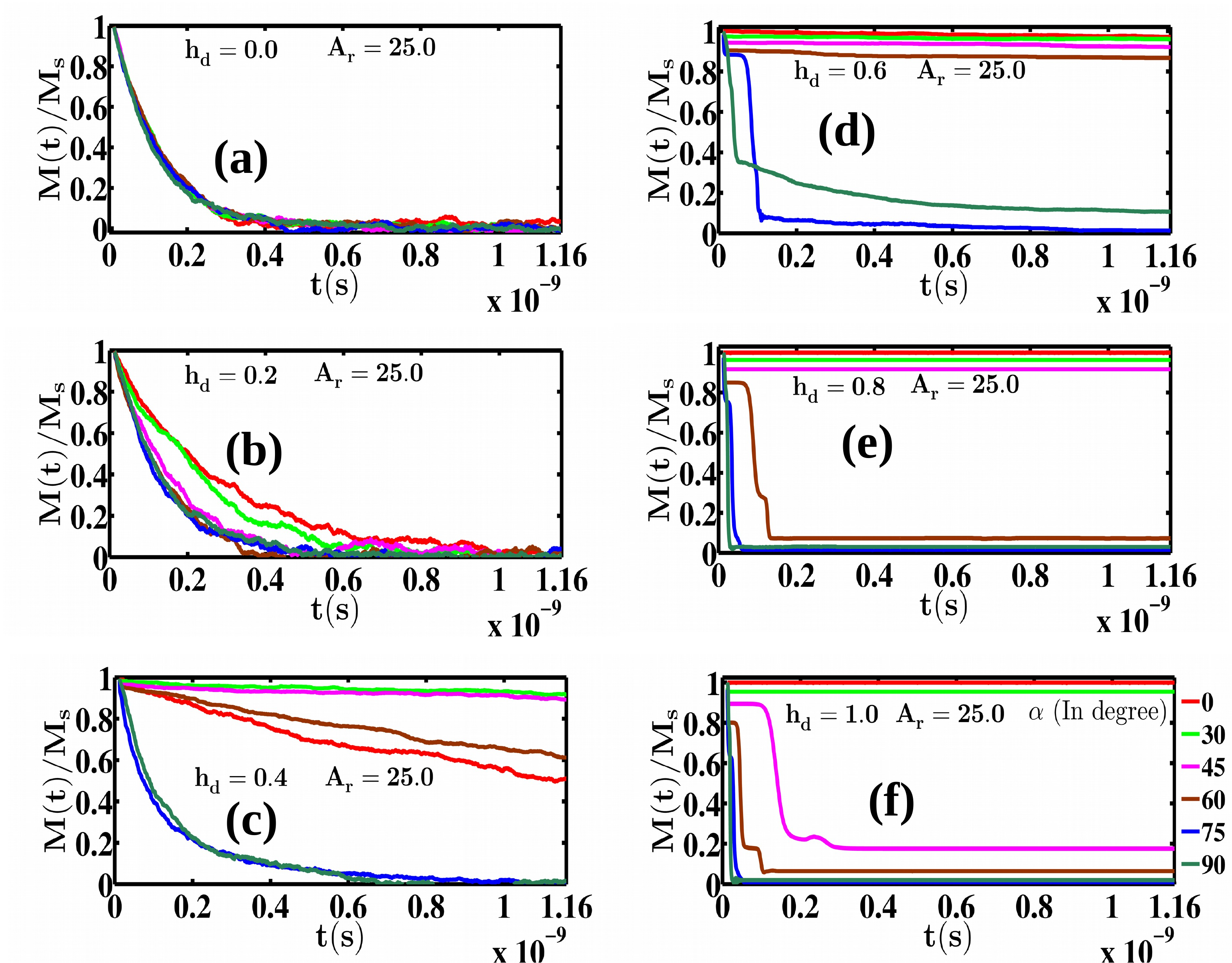}
\caption{Magnetization-decay curves as a function of anisotropy axes orientation with $A^{}_r=25.0$. We have considered six representative values of $h^{}_d$: (a) $h^{}_d= 0.0$, (b) $h^{}_d= 0.2$, (c) $h^{}_d= 0.4$, (d) $h^{}_d= 0.6$, (e) $h^{}_d= 0.8$, and (f) $h^{}_d= 1.0$. Relaxation characteristics is greatly affected by $\alpha$ even for small $h^{}_d$. Magnetization decays very rapidly as $\alpha$ is varied from 0 to $90^\circ$. Remarkably, magnetization ceases to relax with perfectly aligned anisotropy ($\alpha=0^\circ$) and large $h^{}_d$ because of enhancement in ferromagnetic coupling. There is also a fastening in magnetic relaxation for $\alpha>60^\circ$.}
\label{figure4}
\end{figure}

\begin{figure}[!htb]
\centering\includegraphics[scale=0.40]{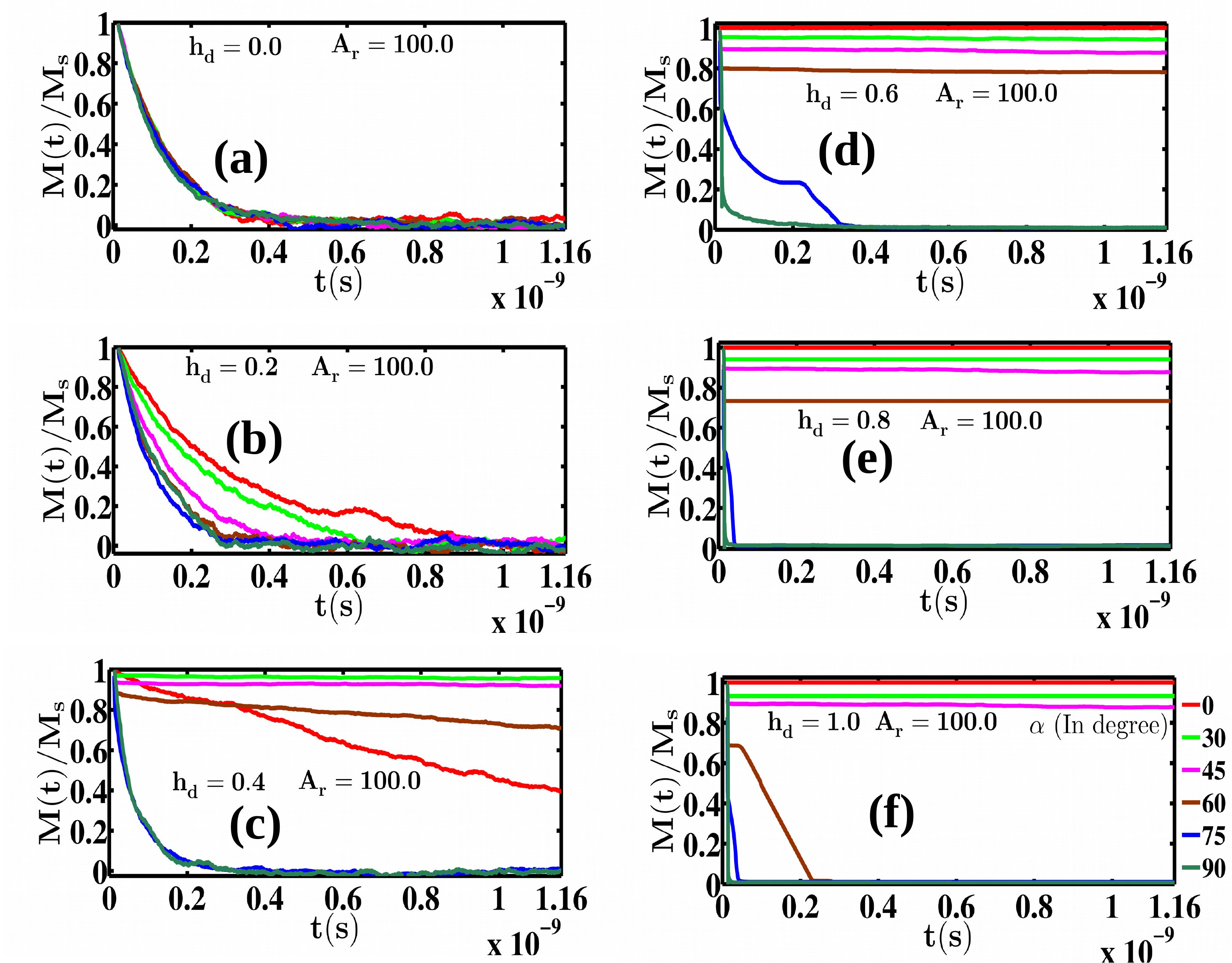}
\caption{Magnetization relaxation curves for various of $h^{}_d$ and $\alpha$ with very large aspect ratio $A^{}_r=100$. We have considered six representative values of $h^{}_d$: (a) $h^{}_d= 0.0$, (b) $h^{}_d= 0.2$, (c) $h^{}_d= 0.4$, (d) $h^{}_d= 0.6$, (e) $h^{}_d= 0.8$, and (f) $h^{}_d= 1.0$. There is a fastening in magnetization-decay with $\alpha$ even in the case of weakly interacting MNPs. While for large $h^{}_d$, magnetization relaxes extremely slowly for $\alpha\leq45^\circ$. Interestingly, the magnetization decays extremely rapidly for $\alpha>60^\circ$ in such a case.}
\label{figure5}
\end{figure}

We next study the time evolution of magnetization in the systems with very large aspect ratios. We plot the magnetization-decay $M(t)/M^{}_s$ versus $t$ curves for $A^{}_r=25.0$ and $100$ in Fig.~(\ref{figure4}) and Fig.~(\ref{figure5}), respectively. All other parameters are the same as that of Fig.~(\ref{figure3}). In the absence of dipolar interaction ($h^{}_d=0.0$), the magnetization relaxation curve is perfectly exponentially decaying, similar to that of the square arrangement of MNPs ($A^{}_r=1.0$). The direction of anisotropy axes starts to affect the relaxation properties even with weakly interacting MNPs ($h^{}_d\approx0.2$). In this case, there is a fastening of magnetic relaxation as $\alpha$ is varied from 0 to $90^\circ$. In the presence of moderate dipolar interaction ($h^{}_d\approx0.4$), the magnetization relaxes slowly for $\alpha\leq45^\circ$. There is a fastening in the magnetic relaxation with $\alpha>45^\circ$. In the case of enormous dipolar interaction strength ($h^{}_d\geq0.6$), the magnetization does not relax at all for perfectly aligned anisotropy ($\alpha=0^\circ$); the same is true for $\alpha\leq60^\circ$. The magnetization decays extremely rapidly for $\alpha>60^\circ$ and large dipolar interaction strength $h^{}_d>0.6$.

\begin{figure}[!htb]
\centering\includegraphics[scale=0.40]{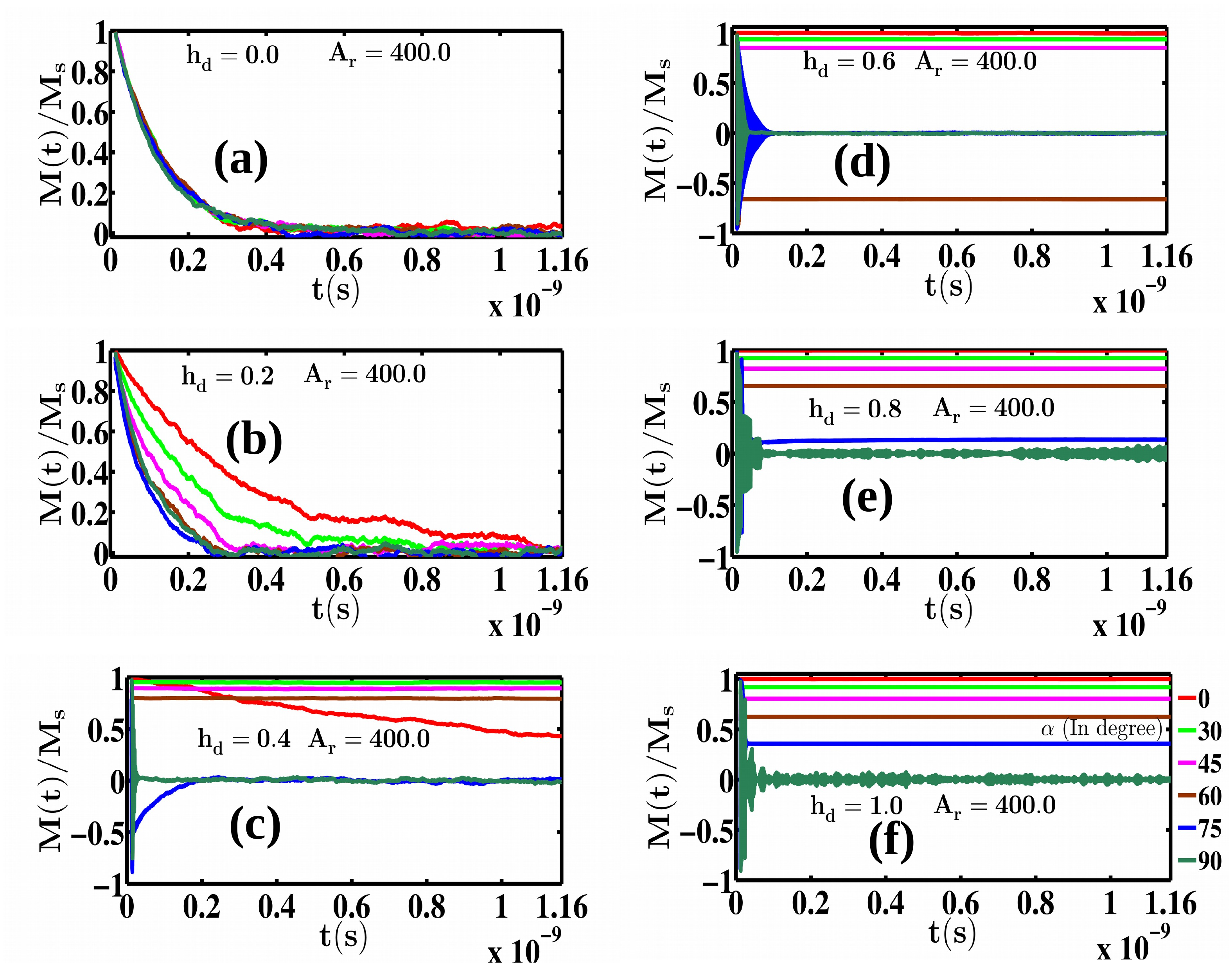}
\caption{Magnetization-decay $M(t)/M^{}_s$ versus $t$ curves for huge $A^{}_r=400$ as  a function of $\alpha$ and $h^{}_d$. We have considered six representative values of $h^{}_d$: (a) $h^{}_d= 0.0$, (b) $h^{}_d= 0.2$, (c) $h^{}_d= 0.4$, (d) $h^{}_d= 0.6$, (e) $h^{}_d= 0.8$, and (f) $h^{}_d= 1.0$. There is a strong effect of $\alpha$ on the rate of magnetization-decay even with a small $h^{}_d=0.2$. There is a rapid decay of magnetization with $\alpha$ in this case. Interestingly, the magnetization does not relax even with moderate values of $h^{}_d$ and $\alpha=0^\circ$. In the presence of moderate dipolar interaction and $\alpha>60^\circ$, there is a complete reversal of magnetization from the saturated situation ($+1$) to $-1$ state as soon as the external field is removed. The anisotropy axis and dipolar field get aligned antiparallel to each other, which helps the magnetization switch its direction just after removing the external magnetic field.}
\label{figure6}
\end{figure}

To understand the effect of anisotropy axis orientation on magnetic relaxation in a system with a huge aspect ratio, we study the time evolution of magnetization for $A^{}_r=400$ in Fig.~(\ref{figure6}); the system corresponds to a one-dimensional array of MNPs. All other parameters are the same as that of Fig.~(\ref{figure5}). The functional form of the magnetization is exponentially decaying for non-interacting MNPs array ($h^{}_d=0.0$), irrespective of $\alpha$ as expected. There is a strong effect of $\alpha$ on the rate of magnetization-decay even with weakly interacting MNPs ($h^{}_d\approx0.2$). There is a fastening of magnetization relaxation as $\alpha$ is varied from $0^\circ$ to $90^\circ$ for $h^{}_d\approx0.2$. Interestingly, the magnetization does not relax even with moderate dipolar interaction strength for perfectly aligned anisotropy axes ($\alpha=0^\circ$). It is because the dipolar interaction promotes ferromagnetic coupling in this case~\cite{anand2019}. There is a rapid decay of magnetization as $\alpha$ is varied from 0 to $90^\circ$. Remarkably, all the magnetic moments of the system change their directions from the saturated state (along $y$-direction) $+1$ to $-1$ (along $-y$-direction) in unison, resulting in a complete reversal of magnetization [$M(t)/M_s=-1$] for $\alpha>60^\circ$. It is due to the fact that the dipolar field and anisotropy axis are antiparallel to each other in such cases~\cite{anand2019}. As a consequence, magnetic moment momentarily reverses their orientations as soon as the external field is removed. {\color{black} The fastening and slowing down of magnetic relaxation with $\alpha$ and $h^{}_d$ is in qualitative agreement with the work of Laslett {\it et al.} and Hovorka {\it et al.}~\cite{laslett2016,hovorka2014}.}
%These results are in perfect aggrement with our recent theoretical and numerical work~\cite{anand2019}.

\begin{figure}[!htb]
	\centering\includegraphics[scale=0.40]{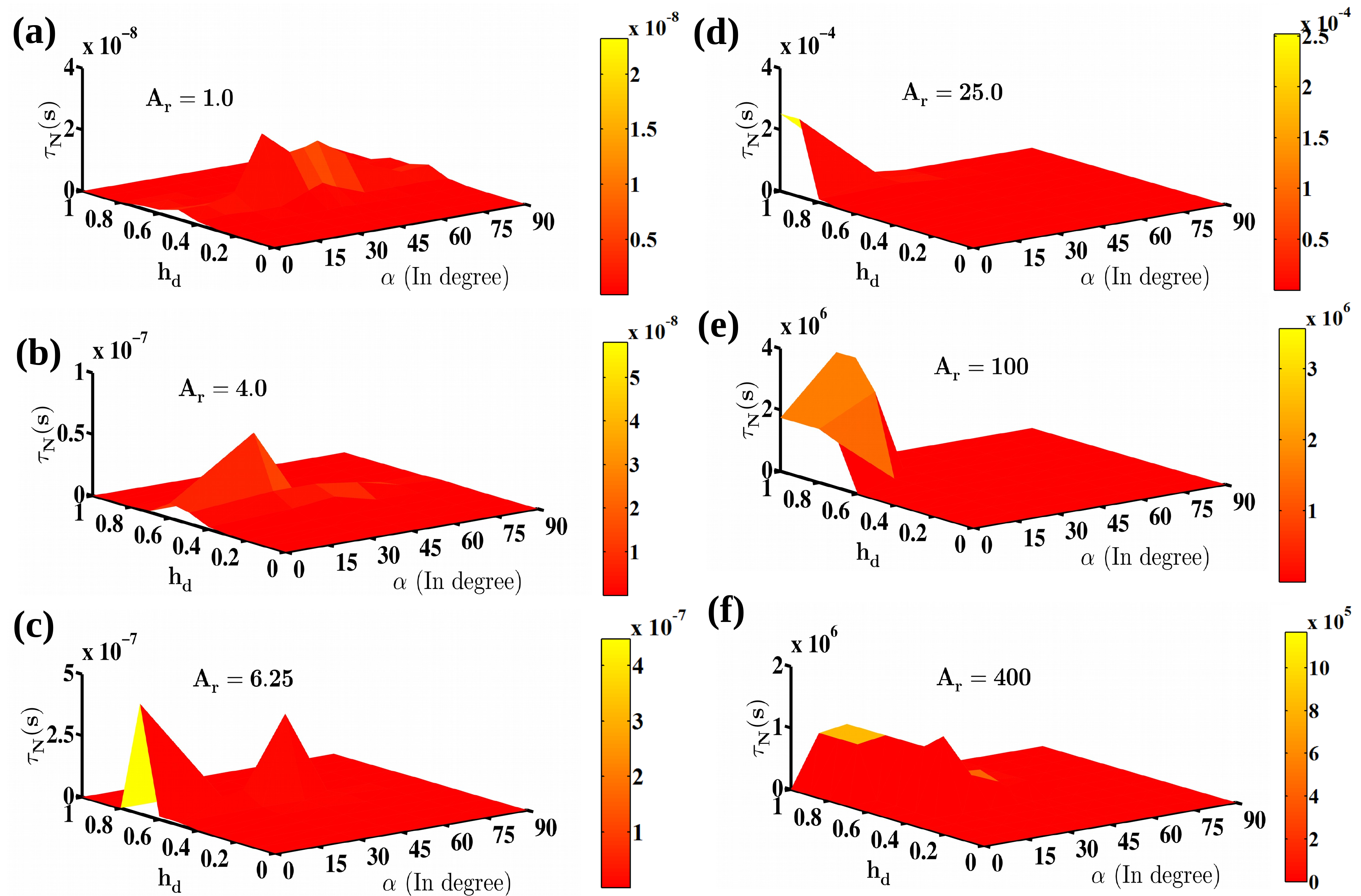}
	\caption{Variation of effective Neel relaxation time $\tau^{}_N$ as a function of $h^{}_d$ and $\alpha$ for various values of $A^{}_r$. We have considered six representative values of $A^{}_r$: (a) $A^{}_r= 1.0$, (b) $A^{}_r= 4.0$, (c) $A^{}_r= 6.25$, (d) $A^{}_r= 25.0$, (e) $A^{}_r= 100$, and (f) $A^{}_r= 400$. There is a weak dependence of $\tau^{}_N$ on $\alpha$ for small $h^{}_d$ and $A^{}_r\leq25.0$. While for large $h^{}_d$, $\tau^{}_N$ decreases with $\alpha$. $\tau^{}_N$ depends strongly on $\alpha$ for huge $A^{}_r$. There is a rapid decrease in $\tau^{}_N$ as $\alpha$ is incremented from 0 to $90^\circ$.}
	\label{figure7}
\end{figure}

Finally, we study the variation of $\tau^{}_N$ as a function of $h^{}_d$ and $\alpha$ in Fig.~(\ref{figure7}). We have varied $\alpha$ between 0 to $90^\circ$ and $h^{}_d$ from 0 to 1.0. We have considered six representative values of aspect ratio $A^{}_r$ of the system. In the presence of weak dipolar interaction ($h^{}_d<0.4$), $\tau^{}_N$ does not depend on the direction of the anisotropy axes, i.e. $\alpha$ with $A^{}_r\leq25.0$. While for large dipolar interaction strength, $\tau^{}_N$ decreases with $\alpha$. $\tau^{}_N$ depends strongly on $\alpha$ in the  highly anisotropic system even with moderate dipolar interaction strength. In the case of perfectly aligned anisotropy axes, $\tau^{}_N$ is the maximum. It is because the strength of ferromagnetic coupling is largest in such a case. There is a rapid decrease in $\tau^{}_N$ as $\alpha$ is incremented from 0 to $90^\circ$. It is due to the fact that the dipolar field tends to get aligned antiparallel to the direction of the anisotropy axis as $\alpha$ is varied from 0 to $90^\circ$. Consequently, magnetization reverses its direction very rapidly as soon as the external magnetic field is switched off. {\color{black} These results can be used in
choosing precise values of aspect ratio, dipolar interaction strength and anisotropy axis orientational angle to obtain the desired relaxation time, which could be useful in digital information storages applications}.

\section{Summary and Conclusion}
Now we summarize the main results presented in this work. In the presence of negligible and small dipolar interaction strength ($h^{}_d\leq0.2$), the functional form of the magnetization-decay curve is perfectly exponentially decaying. The effective N\'eel relaxation time $\tau^{}_N$ evaluated extracted from the simulated relaxation curve is also in perfect agreement with the value obtained using analytical calculation [using Eq.~(\ref{Neel})]. The magnetization relaxation characteristics are found to be independent of anisotropy axes orientation angle $\alpha$ in the system with aspect ratio $A^{}_r\leq6$. In these cases, a transition point is observed at $h^{}_d=0.4$;  time magnetization-decay dynamics gets slower for $\alpha\leq45^\circ$. On the other hand, magnetization relaxes rapidly with $\alpha>45^\circ$. Irrespective of $\alpha$, magnetization decays very rapidly for large dipolar interaction strength $h^{}_d>0.6$. This fastening of magnetization relaxation is due to enhanced antiferromagnetic coupling induced by dipolar interaction. {\color{black} In the case of large dipolar interaction strength, MacIsaac {\it et al.} also observed the dominance of antiferromagnetic coupling in the square array of magnetic moments in the case of large dipolar interaction strength~\cite{macisaac1996}. Our observations are also in perfect qualitative agreement with the work of De'Bell {\it et al.}~\cite{de1997}}. 
%{\color{blue} It can be explained using the fact that dipolar interaction promotes antiferromagnetic coupling between the moments in the square arrangement of MNPs}.  

In a highly anisotropic system, the anisotropy axes orientation strongly affects the magnetic relaxation mechanism even in the presence of small dipolar interaction ($h^{}_d\approx0.2$). We observe fastening of magnetization relaxation as $\alpha$ is incremented from the perfectly aligned case ($\alpha=0^\circ$) to the perpendicular situation ($\alpha=90^\circ$). On the other hand, magnetization does not relax at all with $\alpha\leq60^\circ$ for strongly dipolar interacting MNPs ($h^{}_d>0.6$). The magnetization decays rapidly for $\alpha>60^\circ$. Interesting physics emerges in the case of huge $A^{}_r=400$. Even in the case of weakly dipolar interacting MNPs, there is a rapid decay of magnetization with $\alpha$. Remarkably, magnetization ceases to relax even with moderate dipolar interaction ($h^{}_d=0.4$) and perfectly aligned anisotropy ($\alpha=0^\circ$). It is because dipolar interaction promotes ferromagnetic coupling in such a case. Magnetization decays extremely rapidly as $\alpha$ is varied from 0 to $90^\circ$. Interestingly, there is a complete magnetization reversal from saturation (+1) to -1 state with $\alpha>60^\circ$. In these cases, the anisotropy axes and dipolar field are antiparallel to each other~\cite{anand2019}. Consequently, magnetic moments find it easier to reverse their orientations as soon as the external magnetic field is switched off. The effective N\'eel relaxation time $\tau^{}_N$ is also found to significantly affected by dipolar interaction strength, anisotropy axes orientation and aspect ratio of the system. $\tau^{}_N$ depends weakly on $\alpha$ for small dipolar interaction and $A^{}_r\leq25.0$. On the other hand, it decreases rapidly with $\alpha$ for appreciable $h^{}_d$. In the case of the system with a very high aspect ratio, $\tau^{}_N$ depends strongly on $\alpha$ even with moderate values of $h^{}_d$. In such a case, there is a rapid fall in $\tau^{}_N$ as $\alpha$ is incremented from 0 to $90^\circ$. It is clearly evident that the presence of antiferromagnetic or ferromagnetic interactions depends strongly on the  the angle between the chain axis and the easy axis of the particle, i.e. $\alpha$ in the case of highly anisotropic system (huge $A^{}_r$). The latter plays an crucial role in  fastening or a slowing down of the relaxation.

%There is a slowing down in the magnetization relaxation as $h^{}_d$ is increased from 0 to 0.5, irrespective of $\alpha$. Thereafter, magnetization relaxes faster for $h^{}_d>0.5$. These observations clearly indicate that for small dipolar interaction strength, the anisotropy energy is dominant, which helps in slowing down of relaxation. The antiferromagnetic coupling is dominant for large dipolar interaction strength, resulting in fastening of magnetization relaxation. There is a strong dependance of $\alpha$ on the relaxation characteristics for extremely large $A^{}_r$. There is a slowing down of magnetization relaxation for $\alpha$ 0 to $45^\circ$. Magnetization relaxes fast for $\alpha$ 45 to $90^\circ$. 

In conclusion,  we have analyzed the effect of dipolar interaction, aspect ratio and anisotropy axes orientation on the magnetic relaxation characteristics in the two-dimensional array of magnetic nanoparticles using kinetic Monte Carlo simulation. The assumption of common anisotropy axes provides extra control in manipulating the relaxation properties computationally. {\color{black} Our consideration of aligned anisotropy also provides the freedom to study the change in the nature of dipolar interactions from ferromagnetic to antiferromagnetic by manoeuvring the strength of dipolar interaction and orientation of the easy axis}. There is a strong  effect of these parameters on the magnetization relaxation. Furthermore, the magnetic relaxation characteristics depend strongly on the anisotropy axes orientation for weakly interacting MNPs. Our results are beneficial for applications in data storage and energy storages where such controlled magnetization alignment and desired structural anisotropy is desirable. The observation made in the present article should also help
the experimentalists to manipulate the relaxation characteristics of dipolar interacting self-assembled arrays of MNPs in a more controlled
manner. 
%These results should also be taken into account in the interpretation of the experiments and efficient usage of magnetic hyperthermia.
%Magnetic nanoparticle superstructures with controlled magnetic alignment and desired structural anisotropy hold promise for applications in data storage and energy storage~\cite{jiang2016}. their easy axes [23]. 

%\section*{ACKNOWLEDGMENTS}
%A part of the numerical simulations presented in this work has been performed in the Department of Physics, Indian Institute of Technology (IIT) Delhi. The author is thankful to Prof. Varsha Banerjee for providing the computational facility at IIT Delhi. The author would also like to thank the anonymous reviewers whose comments have greatly improved this manuscript.

%\section*{DATA AVAILABILITY}
%The data that support the findings of this study are available from the corresponding author upon reasonable request.
\bibliographystyle{h-physrev}
\bibliography{ref}
%\end{document}
\newpage

\newpage

\newpage

\newpage

\newpage

\newpage

\newpage

\end{document}